# Superconducting (Li, Fe)OHFeSe film of high quality and high critical parameters

Yulong Huang[1,2,+], Zhongpei Feng [1,2,+], Shunli Ni [1,2,+], Jun Li [4], Wei Hu [1,2], Shaobo Liu [1,2], Yiyuan Mao [1,2], Huaxue Zhou [1], Fang Zhou [1,2], Kui Jin [1,2,3**], Huabing Wang [4], Jie Yuan[1,3**], Xiaoli Dong [1,2,3**] and Zhongxian Zhao[1,2,3**]

[1] Beijing National Laboratory for Condensed Matter Physics and Institute of Physics, Chinese Academy of Sciences, Beijing 100190, China

[2] University of Chinese Academy of Sciences, Beijing 100049, China

[3] Key Laboratory for Vacuum Physics, University of Chinese Academy of Sciences, Beijing 100049, China

[4] Research Institute of Superconductor Electronics, Nanjing University, Nanjing 210093, China

**Abstract** The superconducting film of $(Li_{1-x}Fe_x)OHFeSe$ is reported for the first time. The thin film exhibits a small in-plane crystal mosaic of $0.22^o$, in terms of the FWHM (full-width-at-half-maximum) of x-ray rocking curve, and an excellent out-of-plane orientation by x-ray $\varphi$-scan. Its bulk superconducting transition temperature ($T_c$) of 42.4 K is characterized by both zero electrical resistance and diamagnetization measurements. The upper critical field ($H_{c2}$) is estimated to be 79.5 T and 443 T, respectively, for the magnetic field perpendicular and parallel to the *ab* plane. Moreover, a large critical current density ($J_c$) of a value over 0.5 MA/cm$^2$ is achieved at ~20 K. Such a $(Li_{1-x}Fe_x)OHFeSe$ film is therefore not only important to the fundamental research for understanding the high-$T_c$ mechanism, but also promising in the field of high-$T_c$ superconductivity application, especially in high-performance electronic devices and large scientific facilities such as superconducting accelerator.

High-quality superconducting thin films take an important role in the applications and basic research of high-$T_c$ superconductivity. In both the aspects, iron-based superconductors feature the merit of rich physical phenomena, high superconducting critical parameters (including the transition temperature $T_c$, the upper critical field $H_{c2}$ and the critical current density $J_c$) and small anisotropy [1-12]. Much progress has been made in the synthesis of iron-based superconducting thin films with high performances [6,9,13-19]. Among them, the monolayer film of binary FeSe on a SrTiO$_3$ substrate, showing an energy gap above 65 K, has triggered great interest due to its different electronic structure from the bulk material of FeSe and the highest $T_c$ for the iron-based family to date [8,20-24]. However, the FeSe monolayer samples are very sensitive to the air and the promoted superconductivity fades away quickly once the number of FeSe layers is increased. These drawbacks make it difficult for most measuring techniques to probe the nature of the high-$T_c$ superconductivity and also hamper practical applications. Therefore, it should be put on the agenda to attain a substitute that is compatible with routine experimental measurements and is more suitable for applications.

The newly discovered (Li$_{1-x}$Fe$_x$)OHFeSe (FeSe-11111) superconductor [25], with a comparable $T_c$ and similar electronic structure to the monolayer FeSe, turns out to be a good candidate. However, due to the hydroxyl ion inherent in the compound, it is impossible to obtain (Li$_{1-x}$Fe$_x$)OHFeSe materials, in both bulk and thin film forms, by conventional high-temperature synthesis methods. Most recently, by developing a hydrothermal ion-exchange technique, we have successfully synthesized big and high-quality single crystals of FeSe-11111[26]. Here, we report for the first time the high-quality single-crystalline superconducting film of (Li$_{1-x}$Fe$_x$)OHFeSe, which has been grown on a LaAlO$_3$ (LAO) substrate by a hydrothermal epitaxial method [27]. The high crystalline quality of the film is demonstrated by x-ray diffraction (XRD) results, showing a single (001) orientation with a small crystal mosaic of 0.22° in terms of the FWHM (full-width-at-half-maximum) of the rocking curve and a uniform fourfold symmetry by the $\varphi$ scan of (101) plane. The bulk superconducting transition at $T_c$ of 42.4 K is confirmed by both electrical transport and magnetic measurements.

Based on systematic magnetoresistance measurements, the upper critical field $H_{c2}$ is estimated to be 79.5 T and 443 T, respectively, for the magnetic field perpendicular and parallel to the *ab* plane. The *I-V* (current vs. voltage) results yield a large critical current $J_c$ of over 0.5 MA/cm$^2$ at ~20 K for the FeSe-11111 thin film.

All the XRD experiments were performed at room temperature on a diffractometer (Rigaku SmartLab, 9 kW), equipped with two Ge (220) monochromators. The *dc* magnetic measurements were conducted on a Quantum Design MPMS-XL1 system with a tiny remnant field less than 4 m Oe. Both the electrical resistivity and *I-V* data were collected on a Quantum Design PPMS-9 system using the standard four-probe method.

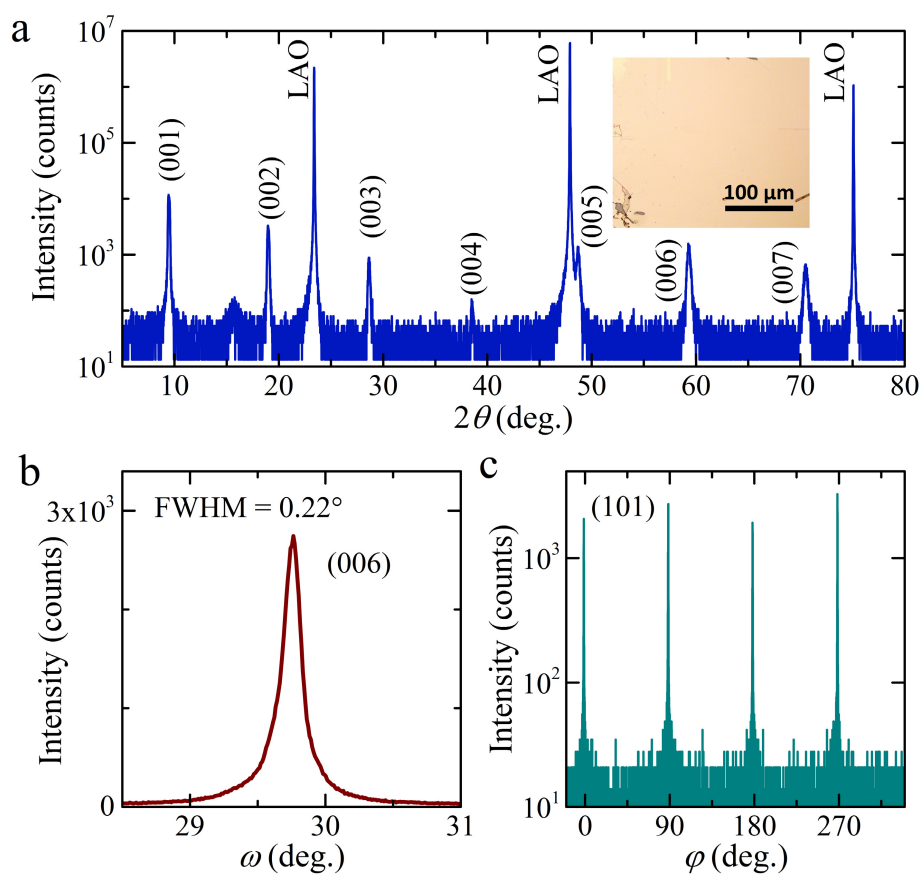

Fig. 1. The XRD characterizations of (Li$_{1-x}$Fe$_x$)OHFeSe film on LaAlO$_3$ (LAO) substrate. (a) The $\theta$-$2\theta$ scan shows only (00*l*) peaks. The inset displays a clean, shiny and mirror-like surface morphology of a cleaved film sample. (b) The rocking curve of (006) reflection with an FWHM = 0.22°. (c) The $\varphi$-scan of the (101) plane. The uniform four-fold symmetry reveals an excellent epitaxial growth.

Fig. 1(a) is a typical XRD pattern of the FeSe-11111 films on LAO substrate. The observation of only (00$l$) reflections indicates its single preferred in-plane orientation. The additional peaks marked with "LAO" are from the substrate. No impurity peaks are detectable. The $c$-axis parameter for FeSe-11111 film calculated from the (00$l$) peaks is 9.329(7) Å, consistent with that for the bulk material[25,26,28]. The inset of Fig. 1(a) displays a clean, shiny and mirror-like surface morphology of a cleaved film sample. Shown in Fig. 1(b) is the double-crystal x-ray rocking curve for the (006) Bragg reflection, with a small FWHM = 0.22°. To our knowledge, this is the best FWHM value obtainable so far among various iron-based superconductors. The $\varphi$-scan of (101) plane in Fig. 1(c) exhibits four successive peaks with an equal interval of 90°, consistent with the $C_4$ symmetry of FeSe-11111 film. That evidences an excellent out-of-plane orientation and epitaxial growth. All the above data demonstrate the high quality of FeSe-11111 film in the aspects of structure, morphology, crystallization and epitaxy, which is even superior to our FeSe-11111 single crystal samples [26]. The superconductivity of FeSe-11111 film is confirmed by the $dc$ diamagnetization, which sets in around 42 K as shown in Fig. 2(a), and by the zero resistivity at 42.4 K as in Fig. 2(b). Despite its high crystalline quality, the full transition width of the film is broad, as commonly seen in iron-based superconducting films[13,16,17,19].

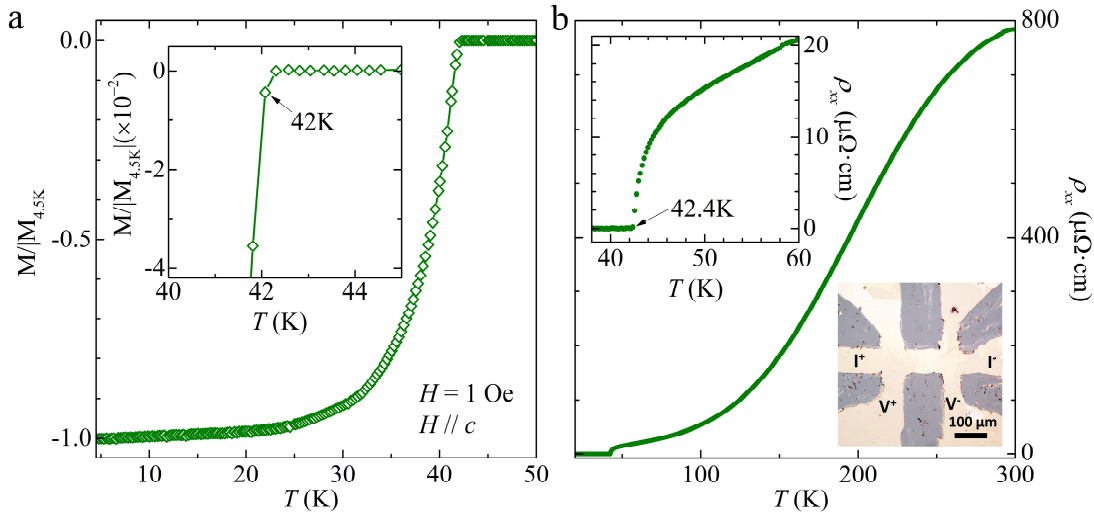

Fig. 2. The superconductivity of $(Li_{1-x}Fe_x)OHFeSe$ film characterized by *dc* magnetization and electrical resistivity. (a) Zero-field-cooling normalized magnetization as a function of temperature. The inset shows the diamagnetization occurring at ~42 K. (b) Temperature dependence of in-plane resistivity. The top inset clearly displays a zero resistivity at 42.4 K. The bottom inset shows the optical image for the microbridge with a width of 80 μm, a length of 120 μm and a thickness of ~100 nm.

In order to obtain both the upper critical field and the critical current density, the cleaved thin film was patterned into a microbridge for electrical transport measurements, as shown in the bottom inset of Fig. 2(b). Since the electrodes of the microbridge are superconducting as well, the heat effects can be considerably eliminated. In particular, the ohmic contact resistance between the film and the silver paste is ~1 ohm, so it has no observable influence on the measurements for a much larger resistance (>100 ohm) of the microbridge. The residual resistivity at zero Kelvin ($\rho_0$) is estimated to be ~10 μΩ·cm by a power law fitting of the $\rho_{xx}$-*T* curve from 180 to 60 K, which is much lower than the previous report on the bulk crystal [26]. Meanwhile, the ratio of the room temperature resistivity to the residual resistivity is much bigger, i.e. RRR = $\rho_{xx}$(300 K)/$\rho_0$ = 78, pointing to a lower impurity scattering or localization in our FeSe-11111 film.

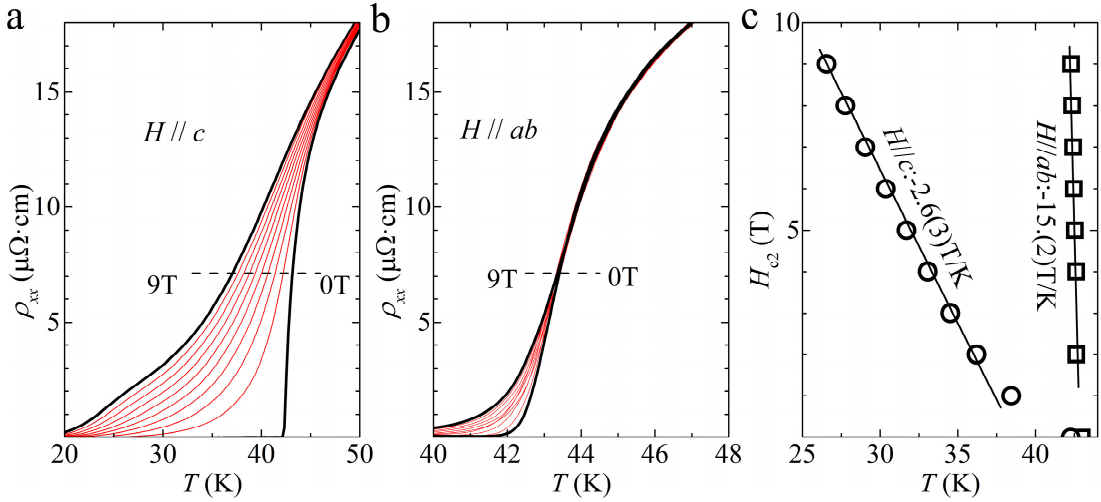

Fig. 3. Magnetoresistance $\rho_{xx}(T, H)$ and upper critical field $H_{C2}(T)$ of $(Li_{1-x}Fe_x)OHFeSe$ film. The in-plane resistivity $\rho_{xx}$ vs. *T* for various magnetic fields along *c*-axis (a) and *ab* plane (b), respectively. (c) Temperature dependence of $H_{c2}(T)$ along the *c*-axis (circle) and within the *ab*-plane (square).

The temperature-dependent upper critical field was obtained from magnetotransport measurements by sweeping the temperature in magnetic fields applied along both the *c*-axis (Fig. 3(a)) and the *ab*-plane (Fig. 3(b)), respectively. The magnetoresistivity properties of the present superconducting thin film are consistent with those of our previous bulk crystals[26]. As shown in Fig. 3(c), the $H_{c2}$ (0 K) values are estimated to be 79.5 T for H // *c* and 443 T for H // *ab* from the Werthamer-Helfand-Hohenberg (WHH) formula, i.e. $H_{c2}(0) = -0.69 T_c dH_{c2}/dT$ with $-dH_{c2}/dT$ the maximum slope in the vicinity of $T_c$. Here, the $H_{c2}$ at limited temperatures is obtained by taking a criterion of the field at 50% of the normal-state resistivity. It is interesting that the $H_{c2}$ along *c*-axis ($H_{c2}^c$) is nearly the same as the value of our single crystal, while the *ab*-plane $H_{c2}^{ab}$ is much higher in the film than the estimated 313 T in the single crystal[26]. The anisotropy $\gamma = H_{c2}^{ab}/H_{c2}^c$ is about 5.6. Such a high upper critical field with a moderate anisotropy is rare in iron-based superconductors[9,11], implying that such a film is well suitable for practical applications of high magnetic fields.

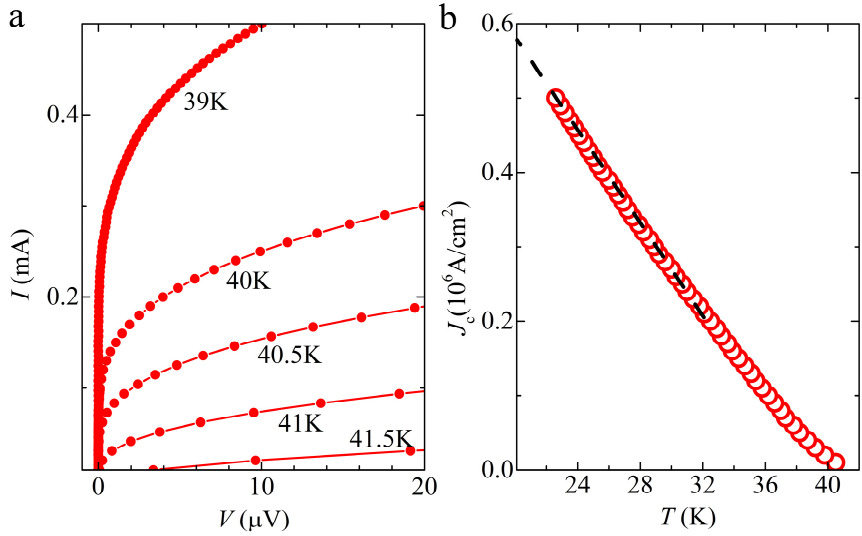

Fig. 4. The *I-V* characteristics and the critical current density $J_c(T)$ of $(Li_{1-x}Fe_x)OHFeSe$ film. (a) The zoom-in view of *I-V* curves near $T_c$. (b) The temperature dependence of $J_c$.

The *I-V* characteristics of FeSe-11111 film were investigated on a narrow bridge with a thickness of 20 nm and a width of 50 μm, shown in Fig. 4(a). The critical current density $J_c$ of the film was extracted from the temperature-dependent *I-V* curves, and is plotted as a function of temperature (Fig. 4(b)). Here, we take a commonly used criterion, 1 μV cm$^{-1}$, as the destruction of the superconducting transportation. It should be mentioned that the critical current already exceeds the upper limit of the allowed current (5 mA) in the PPMS system at 22 K, which means a pretty large critical current density $J_c$ (> 0.5 MA/cm$^2$).

In summary, the superconducting (Li$_{1-x}$Fe$_x$)OHFeSe film is reported for the first time. The high crystalline quality of the thin film is demonstrated by a small in-plane crystal mosaicity of 0.22º and an excellent out-of-plane orientation. Importantly, the (Li$_{1-x}$Fe$_x$)OHFeSe film exhibits high superconducting critical parameters, including the bulk transition temperature $T_c$ of 42.4 K, the large critical current density $J_c$ of over 0.5 MA/cm$^2$ at ~20 K, and the upper critical field $H_{c2}$ at zero temperature of 79.5 T for *H // c* and 443 T for *H // ab* plane, which are among the highest values reported so far for iron-based superconductors. Our results indicate that the (Li$_{1-x}$Fe$_x$)OHFeSe film is promising for superconductivity applications, for instance, in high-performance filters, superconducting cavity resonators and accelerators.

**Acknowlegement** This work is supported by the National Basic Research Program of China (Grant No. 2017YFA0303000), the National Science Foundation of China (Grant Nos. 11574370, 11234006, & 61501220), the Strategic Priority Research Program and Key Research Program of Frontier Sciences of the Chinese Academy of Sciences (Grant Nos. QYZDY-SSW-SLH001, QYZDY-SSW-SLH008 and XDB07020100).

[+] These authors contribute equally
[**] To whom correspondence should be addressed. E-mail: zhxzhao@iphy.ac.cn; dong@iphy.ac.cn; yuanjie@iphy.ac.cn; kuijin@iphy.ac.cn